\documentclass[aps,pof,reprint,10pt,superscriptaddress,showpacs]{revtex4-1}
\usepackage{amsfonts} 
\usepackage{amsmath}
\usepackage{amssymb}
\usepackage{graphicx}%\[  \]
\usepackage{subfigure}
\usepackage{color}
\usepackage{color}
\usepackage{soul}
\usepackage{cancel}
\usepackage{ulem}
\newcommand*\xbar[1]{%
  \hbox{%
    \vbox{%
      \hrule height 0.5pt % The actual bar
      \kern0.5ex%         % Distance between bar and symbol
      \hbox{%
        \kern-0.1em%      % Shortening on the left side
        \ensuremath{#1}%
        \kern-0.1em%      % Shortening on the right side
      }%
    }%
  }%
} 
\begin{document}
\title{Modulation of electromagnetic waves  in a relativistic degenerate plasma at finite temperature}
\author{Sima Roy}
\email{simaroy031994@gmail.com}
\affiliation{Department of Mathematics, Siksha Bhavana, Visva-Bharati University, Santiniketan-731 235, West Bengal, India}
\author{A. P. Misra}
\homepage{Author to whom correspondence should be addressed}
\email{apmisra@visva-bharati.ac.in}
\affiliation{Department of Mathematics, Siksha Bhavana, Visva-Bharati University, Santiniketan-731 235, India}
\author{A. Abdikian}
\email{abdykian@gmail.com}
\affiliation{Department of Physics, Malayer University, Malayer 65719-95863, Iran}
\begin{abstract}
We study the modulational instability (MI) of a linearly polarized electromagnetic (EM) wave envelope in an intermediate regime of relativistic degenerate plasmas at a finite temperature $(T\neq0)$ where the thermal energy $(K_BT)$ and the rest-mass energy $(m_ec^2)$ of electrons do not differ significantly, i.e.,  $\beta_e\equiv K_{B}T/m_{e}c^2\lesssim~(\rm{or}~\gtrsim) 1$, but, the Fermi energy $(K_BT_F)$ and the chemical potential energy $(\mu_e)$ of electrons are still a bit higher than the thermal energy, i.e., $T_F>T$ and  $\xi_{e}=\mu_e/K_{B}T\gtrsim1$. 
Starting from a set of relativistic fluid equations for degenerate electrons at finite temperature, coupled to the EM wave equation and using the multiple scale perturbation expansion scheme,  a one-dimensional nonlinear Sch{\"o}dinger (NLS) equation is derived, which describes the evolution of slowly varying amplitudes of EM wave envelopes. Then we study the  MI of the latter in two different regimes, namely  $\beta_e<1$ and $\beta_e>1$. Like unmagnetized classical cold plasmas, the modulated EM envelope is always unstable in the region $\beta_e>4$. However, for $\beta_e\lesssim1$ and $1<\beta_e<4$, the wave can be stable or unstable depending on the values of the EM wave frequency, $\omega$  and the parameter $\xi_e$. We also obtain the instability growth rate for the modulated wave and find a significant reduction by increasing the values of either $\beta_e$ or $\xi_e$. Finally, we present the profiles of the traveling EM waves in the form of bright  (envelope pulses) and dark  (voids) solitons, as well as the profiles (other than traveling waves) of the Kuznetsov-Ma breather, the Akhmediev breather, and the Peregrine solitons as EM rogue (freak) waves, and discuss their characteristics in the regimes of $\beta_e\lesssim1$ and $\beta_e>1$.  

\end{abstract}
\maketitle
\section{Introduction} \label{sec-intro}
 A high-power laser pulse propagating through plasmas causes many relativistic and nonlinear absorbing effects. Among the varieties of nonlinear phenomena, the interaction of relativistic electromagnetic (EM) waves with plasmas was first investigated by Akhiezer and Polovin in  $1956$ \cite{akhiezer1956}.  They used the coupled Maxwell and relativistic electron fluid equations for modeling the interaction of intense EM waves with plasmas and obtained exact nonlinear wave solutions for describing the propagation of intense laser pulses in plasmas.  Such interactions can also result in other nonlinear phenomena, including the self-focusing \cite{esarey1997}, the harmonic generation \cite{mori1993, shen1995}, the transition from wakefield generation to soliton formation \cite{holkundkar2018, roy2019}, and the generation of large amplitude plasma waves \cite{umstadter2003, sprangle1990}. However, among those, the most interesting phenomenon is the formation of relativistic EM solitons. The latter are localized structures that are self-trapped by a locally modified plasma refractive index due to an increase in the relativistic electron mass and a drop in the electron plasma density by the EM wave-driven ponderomotive force.        
\par 
Several authors have focused their attention to study the generation of EM solitons in plasmas. Kozlov \textit{et al}. \cite{kozlov1979} studied the EM solitons of circularly polarized EM waves in cold plasmas with the effects of the relativistic and striction nonlinearities. By multidimensional particle-in-cell (PIC) simulations, Bulanov \textit{et al}. \cite{bulanov1999, bulanov2001} reported the generation of relativistic solitons in laser-plasma interactions. Recently, the existence and stability of linearly \cite{roy2020} as well as circularly \cite{roy2022} polarized EM solitons were studied in the framework of a generalized nonlinear Schr{\"o}dinger (NLS) equation in relativistic degenerate dense plasmas using the well known Vakhitov-Kolokolov criterion by Roy \textit{et. al.} \cite{roy2020,roy2022} where stationary EM soliton solutions were shown analytically to be stable, in agreement with the model simulation. 
%%%%%%%%%%%%%%%%%%%%%%%
%%%%%%%%%%%%%%%%%%%%%%%%%  
\par 
Modulational instability (MI) is one of the most paramount phenomena in the nonlinear wave theory \cite{liao2023, douanla2022}. It occurs due to the interplay between the nonlinearity and dispersion/diffraction effects in the medium. It is an efficient mechanism for the occurrence of some other nonlinear phenomena such as envelope solitons (bright and dark) \cite{sultana2011,arriaga2015}, envelope shocks \cite{sultana2012}, and freak (or rogue) waves \cite{mckerr2014}. Benjamin and Feir \cite{benjamin1967} theoretically and experimentally established this phenomenon for hydrodynamic waves. Ostrovsky \cite{ostrovsky1967}  studied the self-modulation of nonlinear electromagnetic waves owing to its application to waves in nonlinear media with cubic nonlinearity.
Later, various aspects of the nonlinear propagation of electromagnetic waves  and the modulational instability of electromagnetic solitons  were studied in relativistic magnetized and unmagnetized plasmas (See, e.g., Refs. \cite{tsintsadze1979,stenflo1979,shukla1984,shukla1986}).  In Ref. \cite{tsintsadze1979}, the authors have shown that the relativistic mass variation of electrons can have important effects on the modulational instability of small amplitude waves. In another work, Stenflo \textit{et al.} \cite{stenflo1979} showed that in laser-plasma interactions, new types of circularly polarized waves can appear which undergo modulational instability.   
%%%%
Recently, Rostampooran \textit{et. al}. \cite{rostampooran2017} have investigated the circularly polarized intense EM wave propagating in a weakly relativistic plasma using the mixed Cairns-Tsallis distribution function where the ions are assumed to be stationary and showed that rising of the density of nonthermal electrons increases the amplitude of solitons and rising of nonextensive electrons decreases the amplitude of solitons. In other work \cite{borhanian2009}, Borhanian \textit{et al.} have shown the existence of bright envelope solitons in the nonlinear propagation of extra-ordinary waves in a magnetized cold plasma. They showed that the bright soliton broadens when the wave frequency increases from the near critical frequency, and its width decreases for larger values of the carrier wave frequency. They also noted that a bright envelope soliton for the fast mode represents the possible stationary solutions of the nonlinear Schr{\"o}dinger equation and nonlinear coupling of circularly polarized EM waves with the background plasma. 
\par 
In this paper, we aim to advance the theory of MI of a linearly polarized EM wave propagating in an unmagnetized relativistic degenerate dense plasma at finite temperature and focus on an intermediate regime where the thermal energy and the rest-mass energy of electrons do not differ significantly, but, the Fermi energy and the chemical potential energy of electrons are still higher than their thermal energy. In this way, the present model somewhat advances the work of Borhanian \textit{et. al.} \cite{borhanian2009} but in an unmagnetized plasma with the effects of the finite-temperature degenerate pressure. The latter significantly modifies the stability and instability domains, not reported before in the literature. Starting from a one-dimensional relativistic fluid model coupled to the EM wave equation, we develop a   multiple-scale expansion scheme to derive the NLS equation, which governs the existence of traveling wave solutions (e.g., bright and dark-type envelope solitons) as well as breather-types of solutions (non-traveling wave) as EM rogue waves \cite{liu2023, slunyaev2021}.

\section{Basic Equations} \label{sec-basic}
 We consider the nonlinear propagation of linearly polarized EM waves in an unmagnetized plasma  with relativistic flow of degenerate electrons at finite temperature and immobile positive ions. We assume that the finite amplitude linearly polarized EM waves propagate in the $z$-direction, i.e., all the dynamical variables vary with the space coordinate $z$ and time variable $t$.  The Coulomb gauge condition gives the parallel and perpendicular (to $z$) components of the wave electric field as $E_{z}=-\partial \phi/\partial z$ and   $\mathbf{E_{\perp}}=-\partial \mathbf{A}/\partial t$, where $\phi$ and ${\bf A}$, respectively, denote the scalar and the vector potentials. The EM wave equation together with the relativistic fluid equations for degenerate electrons are \cite{holkundkar2018,misra2018}  
\begin{equation} \label{eq-emw}
\frac{\partial^2 \mathbf{A} }{\partial z^2}-\frac{1}{c^2}\frac{\partial ^2 \mathbf{A}}{\partial t^2}=\frac{e^2c^2n_{e}n_{0}\mathbf{A}}{\epsilon_{0}H_{0}},
\end{equation}
\begin{equation} \label{eq-cont}
\frac{\partial n_{e}}{\partial t}+ \frac{\partial}{\partial z}(n_{e}v_{z})=0,
\end{equation}
\begin{equation} \label{eq-moment}
\begin{split}
\frac{dv_{z}}{dt}=&\frac{en_{0}c^2}{H_{0}\gamma}\left(1-\frac{H_{0}}{m_{e}c^2n_{0}}\frac{v_{z}^2}{c^2}\right)\frac{\partial \phi}{\partial z}-\\& \frac{e^2n_{0}^2c^2}{2H_{0}^2\gamma^2}\left(\frac{\partial \mathbf{A^2}}{\partial z}+\frac{H_{0}^2}{n_{0}^2m_{e}^2c^{4}}\frac{v_{z}}{c^2}\frac{\partial \mathbf{A^2}}{\partial t}\right)- \frac{n_{0}c^2}{H_{0}\gamma^2 n}\frac{\partial P_{e}}{\partial z},
\end{split}
\end{equation}
\begin{equation} \label{eq-poisson}
\frac{\partial^2 \phi}{\partial z^2}=\frac{e\gamma}{\epsilon_{0}}(n_{e}-n_{0}),
\end{equation}
where $e,~ m_{e}$, and $n_{e}$ are, respectively, the charge, mass, and number density of electrons  and  $n_0$ is the background number density of electrons or ions. Also, $H_{0}$ is the equilibrium value of the enthalpy per unit volume of the electron fluid,  measured in the rest frame, which involves the relativistic pressure, $P_e$ at $T \neq 0$ K, the rest mass energy density, and the internal energy density \cite{misra2018}.  Furthermore, $v_{z}$ is the parallel component of the electron fluid velocity, $c$ is the speed of light in vacuum,    $d/dt \equiv \partial/\partial t + v_{z}\partial/\partial z$, and $\gamma$ is the Lorentz factor, given by,
\begin{equation} \label{eq-gamma}
\gamma=\sqrt{\frac{1+a^2}{1-V^2}},
\end{equation}
where ${\bf a}=e \mathbf{A}/m_ec^2$ and $V=v_{z}/c$.
\par  Next,  we normalize the physical quantities  according to  $\phi\rightarrow e\phi/m_ec^2$,  $t \rightarrow \omega_p t$, $z \rightarrow z/\lambda_D$, and  $N=\gamma n_{e}/n_{c}$  with $n_{c}=\epsilon_{0}\omega^2H_{0}/n_{0}e^2c^2$ and $\alpha=m_{e}n_{0}c^2/H_{0}$, where $\omega_p$ is the electron plasma oscillation frequency and $\lambda_D$ is the electron Debye screening length. Also, the pressure $P_{e}$ is normalized by $n_{0}K_{B}T$.  Thus,  Eqs. \eqref{eq-emw} to \eqref{eq-poisson} reduce to (writing the EM wave equation in a scalar form)
\begin{equation} \label{eq-norm-emw}
\frac{\partial^2 a}{\partial z^2}-\frac{\partial^2 a}{\partial t^2}=\frac{Na}{\gamma},
\end{equation}
\begin{equation} \label{eq-norm-cont}
\frac{\partial N}{\partial t}+\frac{\partial}{\partial z}(NV)=0,
\end{equation}
\begin{equation} \label{eq-norm-moment}
\begin{split}
N\gamma^2\frac{\partial V}{\partial t}= & N\alpha\gamma\left(1-\frac{V^2}{\alpha}\right)\frac{\partial \phi}{\partial z}-N\frac{\alpha^2}{2}\left(\frac{\partial a^2}{\partial z}+\frac{V}{\alpha^2}\frac{\partial a^2}{\partial t}\right) \\ & -\frac{\gamma\alpha}{m_{e}c^2}\frac{\partial P_{e}}{\partial z},
\end{split}
\end{equation}
\begin{equation} \label{eq-norm-poisson} 
\frac{\partial^2 \phi}{\partial z^2}=\frac{1}{\alpha}(N-N_{0}).
\end{equation}
\par
Using the Fermi-Dirac statistics, the expression for the relativistic pressure at finite temperature ($T\neq 0$ K) can be obtained as \cite{boshkayev2016,dey2023}
\begin{equation} \label{eq-pressure}
P_{e}=\frac{2^{3/2}}{3\pi^2 \hbar^3}m_{e}^4c^5\beta_{e}^{5/2} \left[ F_{3/2}(\eta,\beta_{e})+\frac{\beta_{e}}{2}F_{5/2}(\eta,\beta_{e}) \right],
\end{equation}
where
\begin{equation} \label{eq-int-F}
F_{k}(\eta,\beta_{e})=\int_{0}^{\infty}\frac{t^{k}\sqrt{1+(\beta_{e}/2)t}}{1+\exp({t-\eta})}dt,
\end{equation}
is the relativistic Fermi-Dirac integral in which $\beta_{e}=K_{B}T/m_{e}c^2$ is the relativistic parameter, $t=E(p)/K_{B}T$ with $E(p)=\sqrt{c^2p^2+m_{e}^2c^4}-m_ec^2$ denoting the relativistic energy  and $p$ the electron momentum, and $\eta=(\mu_{e}+e\phi)/(K_{B}T)$ is the normalized electrochemical potential energy. 
\par 
An explicit expression of the pressure $P_e$ in terms of $\eta$ and $\beta_e$ is much complicated to obtain. Also, its expressions in the two extreme limits, i.e., the non-relativistic or weakly relativistic  $(\beta_{e}\equiv K_{B}T/m_{e}c^2 \ll 1)$ and the ultra-relativistic $(\beta_{e} \gg1)$ regimes of Fermi gas have been considered before in different contexts. We are, however,  interested in an intermediate regime in which the electron thermal energy and the rest mass energy do not differ significantly, i.e., $\beta_e\sim1$,  i.e., either $\beta_{e}\lesssim1$ or $\beta_{e}\gtrsim1$. From Eqs. \eqref{eq-pressure} one can obtain the following expressions for the degenerate pressure of electrons in these two different cases  \cite{dey2023}. The strictest case with $\beta_e=1$ is not of interest to the present study.
%%%%%%%%%%
%%%%%
%%%%%%
\begin{widetext}
\begin{equation}
P_{e}=
\begin{cases}
\begin{split}
&\frac{2^{3/2}}{3\pi^2\hbar^{3}}m_{e}^{4}c^{5}\beta_{e}^{5/2}\left[\Bigl\{ \frac{2}{5}(\xi_{e}+\phi)^{5/2} + \frac{\pi^2}{4}(\xi_{e}+\phi)^{1/2}-\frac{7\pi^{4}}{960}(\xi_{e}+\phi)^{-3/2}\Bigr\}+ \right.\\&\left. \frac{\beta_{e}}{2}\Bigl\{\frac{2}{7}(\xi_{e}+\phi)^{7/2}+\frac{5\pi^2}{12}(\xi_{e}+\phi)^{3/2}+\frac{7\pi^4}{192}(\xi_{e}+\phi)^{-1/2}\Bigr\}\right] & \text{for $\beta_{e}<1$},\\
&\frac{1}{9\pi^2\hbar^{3}}m_{e}^{4}c^{5}\beta_{e}^{7/2}\left[2\Bigl\{(\xi_{e}+\phi)^{3}+\pi^{2}(\xi_{e}+\phi)\Bigr\}+ \right.\\&\left. 3\beta_{e}\Bigl\{\frac{1}{4}(\xi_{e}+\phi)^{4}+\frac{\pi^2}{2}(\xi_{e}+\phi)^{2}+\frac{7\pi^4}{60}\Bigr\}\right] & \text{for $\beta_{e}>1$},
\end{split}
\end{cases}
\end{equation}
\end{widetext}
where
$\xi_{e}=\mu_{e}/K_{B}T$ is the degeneracy parameter for electrons at equilibrium, which satisfies the following  condition at zero relativistic and electrostatic potential energies, i.e.,
\begin{equation}
\sum[1+\exp(\xi_{e})]^{-1} \leq 1,
\end{equation}
and we have assumed $\eta>1$ without loss of generality.
\par
The expression for the pressure gradient in Eq. \eqref{eq-norm-moment} can be written as 
\begin{equation} \label{eq-expan-pressure}
\frac{\partial P_{e}}{\partial z}=\left(\frac{dP_{e}/d\phi}{dn_{e}/d\phi}\right)\frac{\partial n_{e}}{\partial z}\equiv\left(\frac{dP_{e}}{dn_{e}}\right)_{\phi=0}\frac{\partial n_{e}}{\partial z}=\tilde{c_{s}}^2\frac{\partial n_{e}}{\partial z},
\end{equation}
where
\begin{equation} \label{eq-cs}
\tilde{c_{s}}^2=\frac{1}{c^2}\left[\frac{1}{m_{e}}\left(\frac{dP_{e}}{dn_{e}}\right)_{\phi=0}\right]=\nu_{e}K_{B}T/m_{e}c^2=\nu_{e}\beta_{e} 
\end{equation}
is the modified acoustic speed in which 
\begin{equation} \label{eq-nu}
\nu_{e}=\frac23\frac{\xi_{e}A_{e}}{B_{e}}.
\end{equation}
Here, the expressions for   $A_{e}$ and $B_{e}$ are given by
\begin{equation} \label{eq-expan-Ae}
A_{e}=
\begin{cases}
\begin{split}
&\left[\left(1+\frac{\pi^2}{8}\xi_{e}^{-2}+\frac{7\pi^4}{640}\xi_{e}^{-4}\right)+\right.\\&\left.\frac{\xi_{e}\beta_{e}}{2}\left(1+\frac{5\pi^2}{8}\xi_{e}^{-2}-\frac{7\pi^4}{384}\xi_{e}^{-4}\right)\right], &\text{for $\beta_{e}<1$},\\
&\left[\left(1+\frac{\pi^2}{3}\xi_{e}^{-2}\right)+\frac{\xi_{e}\beta_{e}}{2}(1+\pi^2\xi_{e}^{-2})\right], &\text{for $\beta_{e}>1$},
\end{split}
\end{cases}
\end{equation}
\begin{equation} \label{eq-expan-Be}
B_{e}=
\begin{cases}
\begin{split}
&\left[\left(1-\frac{\pi^2}{24}\xi_{e}^{-2}-\frac{7\pi^4}{384}\xi_{e}^{-4}\right)+\right.\\&\left.\xi_{e}\beta_{e}\left(1+\frac{\pi^2}{8}\xi_{e}^{-2}+\frac{7\pi^4}{640}\xi_{e}^{-4}\right)\right], & \text{for $\beta_{e}<1$},\\
&\left[1+\beta_{e}\xi_{e}\left(1+\frac{\pi^2}{3}\xi_{e}^{-2}\right)\right], & \text{for $\beta_{e}>1$}.
\end{split}
\end{cases}
\end{equation}
Next, with the use of Eqs. \eqref{eq-pressure}, \eqref{eq-expan-pressure} and \eqref{eq-cs};  Eq. \eqref{eq-norm-moment} reduces to
\begin{equation} \label{eq-reduced-moment}
\begin{split}
N\gamma^2\frac{\partial V}{\partial t}= & N\alpha\gamma\left(1-\frac{V^2}{\alpha}\right)\frac{\partial \phi}{\partial z}\\ & - N\frac{\alpha^2}{2}\left(\frac{\partial a^2}{\partial z}+\frac{V}{\alpha^2}\frac{\partial a^2}{\partial t}\right)-\gamma\alpha\tilde{c_{s}}^2\frac{\partial N}{\partial z}.
\end{split}
\end{equation} 
Equations \eqref{eq-norm-emw}, \eqref{eq-norm-cont}, \eqref{eq-norm-poisson}, and \eqref{eq-reduced-moment} form the desired set of equations for the propagation of EM waves in an unmagnetized relativistic degenerate plasma at finite temperature.  
\section{Physical Regimes} \label{sec-phy-regime}
In the preceding section \ref{sec-basic}, we have stated the model equations for the nonlinear interactions between linearly polarized EM waves and relativistic degenerate plasmas at a finite temperature. Here, we discuss the physical regimes in which the model equations can be valid and  identify the key parameters and their domains for the nonlinear modulation of EM waves and their evolution as wave envelopes. Clearly, the key parameters are the normalized chemical potential $\xi_e$ and the normalized thermal energy $\beta_e$.  We are mainly interested in an intermediate regime where the electron thermal energy and the rest-mass energy do not differ significantly, i.e., $\beta_e\equiv K_BT/m_ec^2\sim1$, or, more precisely, $\beta_e$ is slightly smaller or larger than unity, i.e.,  $\beta_e<1$  or $\beta_e>1$.    Also, the electrons have energy states between the thermal energy $K_BT$ and the Fermi energy $K_BT_F$ such that $T_F>T$. The latter enforces us to assume the normalized chemical potential $\xi_e$ to be positive and in addition $\xi_e>1$ (since in the derivation of the expression for the Fermi pressure $P_e$ in terms of $\xi_e$ and $\phi$, we have assumed $\eta>1$, which at $\phi=0$ gives $\xi_e>1$). We note that the weakly relativistic and ultra-relativistic plasma regimes can be recovered from Eq. \eqref{eq-pressure} in the two extreme conditions $\beta_e\ll1$ and $\beta_e\gg1$ respectively.  
 Also, the regimes $T_F\gg T$ and $T_F\ll T$, respectively,  correspond to the completely degenerate and nondegenerate plasmas. However, these particular cases are not of interest to the present study. For a fully degenerate plasma, the chemical energy $\mu_e$ may be taken to be approximately the
Fermi energy $K_BT_F$. However, for plasmas with finite temperature degeneracy, $\mu_e$  rather
depends on the temperature $T$ \cite{shi2014}.  As a result, the model equations do not involve the Fermi energy explicitly.  
\par 
    On the other hand, it has been shown that \cite{shi2014}, the values of $\xi_e$ can vary in the range $0\lesssim\xi_e\lesssim20$ for $0.1\lesssim T~(\rm{ev})~\lesssim1.4$. It has also been found that as the electron temperature drops below $10^7$ K, the electron degeneracy parameter $\xi_e$ assumes values from $10^{-2}$ to a value close to $10$ at a smaller value of $T$ \cite{thomas2020}. Thus, it is reasonable to consider values of $\xi_e$ in the regime $1<\xi_e\lesssim10$. This regime of $\xi_e$ together with the conditions $T_F>T$ and $\beta_e<1$  or $\beta_e>1$, can be relevant in the laser-plasma interaction experiments, e.g., at the National Ignition Facility (NIF) \cite{hurricane2014} with the electron number density, $n_0\gtrsim10^{25}~\rm{cm}^{-3}$.
\section{Derivation of the NLSE } \label{sec-NLSE}
  We study the modulational instability of linearly polarized EM wave envelopes in relativistic plasmas. To this end, we use the multiple-scale expansion technique in which the  stretched  coordinates for space and time are expressed by the Lorentz transformations as
  \begin{equation} \label{eq-stretch}
  \xi=\epsilon\gamma_{v}\left(z-v_{g}t\right),~~\tau=\epsilon^2\gamma_{v}\left(t-v_g z\right),
  \end{equation}
where $\gamma_v=1/\sqrt{1-v_g^2}$ is one another Lorentz factor for the relativistic  dynamics of EM waves in the new coordinate frame of reference,   $v_{g}$ is the  group velocity of wave envelopes, to be determined later, and $\epsilon~ (0<\epsilon<1)$ is a small expansion parameter, which measures the weakness of perturbations. It is to be noted that  several authors have used the Galilean transformation for the relativistic dynamics of EM waves (see, e.g.,  \cite{borhanian2009}). However, while Galilean transformation can be a good assumption for nonrelativistic dynamics of waves,  the Lorentz transformation must be considered for the description of relativistic wave dynamics with relativistic plasma flows. Later, we will see how the Lorentz factor contributes to the wave dispersion and nonlinearity of the NLS equation. 
\par
Next, to expand the dynamical variables, we consider the perturbations for the density,  velocity, and  the scalar and vector potentials in the form of an wave envelope, which has slower space and time variations of its amplitude in comparison with the fast space-time scales of the carrier wave (phase) dynamics. Thus, the dynamical variables are expanded as
\begin{equation} \label{eq-expan-N-V-a-phi}
\begin{aligned}
N=1+\sum_{n=1}^{\infty}\epsilon^{n}\sum_{l=-\infty}^{\infty}N_{l}^{(n)}(\xi,\tau)\exp[il(kz-\omega t)],\\
V=\sum_{n=1}^{\infty}\epsilon^{n}\sum_{l=-\infty}^{\infty}V_{l}^{(n)}(\xi,\tau)\exp[il(kz-\omega t)],\\
a=\sum_{n=1}^{\infty}\epsilon^{n}\sum_{l=-\infty}^{\infty}a_{l}^{(n)}(\xi,\tau)\exp[il(kz-\omega t)],\\
\phi=\sum_{n=1}^{\infty}\epsilon^{n}\sum_{l=-\infty}^{\infty}\phi_{l}^{(n)}(\xi,\tau)\exp[il(kz-\omega t)].
\end{aligned}
\end{equation}
For all the state variables defined above, the reality condition  $S_{-l}^{(n)}=S_{l}^{(n)*}$ must be satisfied. Here, the asterisk denotes the complex conjugate (c.c.) of the corresponding physical quantity.
In what follows, we substitute the expansions from Eq. \eqref{eq-expan-N-V-a-phi} into Eqs. \eqref{eq-norm-emw}, \eqref{eq-norm-cont},  \eqref{eq-norm-poisson}, and \eqref{eq-reduced-moment},   and collect  the terms in different powers of $\epsilon$. The results are given in the following subsections \ref{sec-disp}-\ref{sec-nls}.
\subsection{First order perturbations: Linear dispersion relation} \label{sec-disp}
 For the first harmonic of the first order perturbations with $n=1,~ l=1$, we obtain the following relations. 
\begin{equation} \label{eq-N-V-1}
-\omega N_{1}^{(1)} + kV_{1}^{(1)}=0,
\end{equation}
\begin{equation} \label{eq-phi-N-1}
-k^2 \phi_{1}^{(1)}=\frac{1}{\alpha} N_{1}^{(1)},
\end{equation}
\begin{equation} \label{eq-disp0}
(\omega^2-k^2)a_{1}^{(1)}=a_{1}^{(1)},
\end{equation}
\begin{equation} \label{eq-V-phi-N-1}
\omega V_{1}^{(1)}=k\alpha\left(\phi_{1}^{(1)}-\frac{c_{s}^2}{c^2}N_{1}^{(1)}\right).
\end{equation}
From Eqs. \eqref{eq-N-V-1},  \eqref{eq-phi-N-1}, and \eqref{eq-V-phi-N-1}, we obtain
\begin{equation} \label{eq-n-1-l-1}
N_{1}^{(1)}=V_{1}^{(1)}=\phi_{1}^{(1)}=0,
\end{equation}
while from Eq. \eqref{eq-disp0} we obtain  the following linear dispersion relation for  EM waves in an unmagnetized plasma \cite{holkundkar2018}.
\begin{equation} \label{eq-disp1}
\omega^2=1+k^2.
\end{equation}
 From Eq. \eqref{eq-n-1-l-1}, it is seen that the first order perturbations  for the electron density, parallel velocity and scalar potential vanish. This is expected, since these perturbations are associated with the longitudinal motion of the wave electric field.  
\subsection{Second order perturbations: Compatibility condition and harmonic generation}\label{sec-vg}
For the second order zeroth harmonic modes (with $n=2,~l=0$), we obtain
\begin{equation} \label{eq-n-2-l-0}
N_{0}^{(2)}=a_{0}^{(2)}=\frac{\partial V_{0}^{(1)}}{\partial \xi}=\frac{\partial \phi_{0}^{(1)}}{\partial \xi}=0.
\end{equation}
Also, for the second order first harmonic modes (with $n=2, l=1$) we obtain
\begin{equation} \label{eq-n-2-l-1}
N_{1}^{(2)}=V_{1}^{(2)}=\phi_{1}^{(2)}=0,
\end{equation} 
together with the following compatibility condition.
\begin{equation} \label{eq-vg}
v_{g}=\frac{\partial \omega}{\partial k}=\frac{k}{\omega}=\frac{k}{\sqrt{1+k^2}}.
\end{equation}
Evidently, even though the phase velocity of EM waves can be larger than the speed of light $c$ [see Eq. \eqref{eq-disp1}], the group velocity $v_g$ (in its original dimension) remains smaller than $c$. 
\par 
Proceeding in this way,  for $n=2, l=2$, we obtain the following second order second harmonic wave amplitudes, which are generated due to the self-interactions of the carrier waves.
\begin{equation} \label{eq-N-2}
N_{2}^{(2)}=\frac{2k^2\alpha^2}{4\omega^2-1-4k^2\alpha\tilde{c_{s}}^2}\left(a_{1}^{(1)}\right)^2,
\end{equation}
\begin{equation} \label{eq-V-2}
V_{2}^{(2)}=\frac{2\omega k\alpha^2}{4\omega^2-1-4k^2\alpha\tilde{c_{s}}^2}\left(a_{1}^{(1)}\right)^2,
\end{equation}
\begin{equation} \label{eq-phi-2}
\phi_{2}^{(2)}=-\frac{1}{2}\frac{\alpha}{4\omega^2-1-4k^2\alpha\tilde{c_{s}}^2}\left(a_{1}^{(1)}\right)^2,
\end{equation}
\begin{equation} \label{eq-a-2}
a_{2}^{(2)}=0.
\end{equation}
\subsection{Third order perturbations: The NLS equation} \label{sec-nls}
  Finally, we consider the third order perturbation equations with zeroth and first harmonic modes.    Considering the third order zeroth harmonic modes with $n=3,~ l=0$, we obtain
\begin{equation} \label{eq-phi-0-2}
\phi_{0}^{(2)}=\alpha|a_{1}^{(1)}|^2.
\end{equation}
It shows that  a zeroth harmonic mode of the scalar potential is generated due to the nonlinear self-interactions of the first order vector potentials. 
\par 
Next, considering the perturbation equations with $n=3,~ l=1$, we find that the third order first harmonic wave amplitudes vanish, i.e.,
\begin{equation} \label{eq-n-3-l-1}
N_{1}^{(3)}=V_{1}^{(3)}=\phi_{1}^{(3)}=0,
\end{equation}
and eventually we obtain the following NLS equation for the evolution of the slowly varying amplitude $\psi\equiv a_{1}^{(1)}$ of linearly polarized EM wave envelopes, $a(z,t)\sim a_1^{(1)}(\xi,\tau)\exp[i(kz-\omega t)]+\text{c.c.}$, in a relativistic unmagnetized degenerate plasma at finite temperature.
\begin{equation} \label{eq-NLSE}
i\frac{\partial \psi}{\partial \tau}+P\frac{\partial^2 \psi}{\partial \xi^2}+Q|\psi|^2\psi=0,
\end{equation}
where  the group velocity dispersion coefficient $P$ and the cubic nonlinear (Kerr) coefficient $Q$ are given by
\begin{equation} \label{eq-P}
P\equiv\frac{1}{2}\gamma_v\frac{\partial v_{g}}{\partial k}=\frac{\gamma_{v}}{2\omega},
\end{equation}
and
\begin{equation} \label{eq-Q}
Q=\gamma_{v}\left(\frac{1}{2\omega}-\frac{k^2\alpha^2}{\omega\left(4\omega^2-1-4k^2\alpha\tilde{c_{s}}^2\right)}\right).
\end{equation}
It is interesting to note that the Lorentz factor $\gamma_v$, which several authors did not consider in the context of relativistic wave dynamics (see, e.g., \cite{borhanian2009}), contributes to (and thus modifies) both the dispersion and nonlinear coefficients of the NLS equation \eqref{eq-NLSE}.  
By disregarding the relativistic degeneracy pressure proportional to $\tilde{c_{s}}^2$, one can recover the same expressions for $P$ and $Q$ as in Ref. \cite{borhanian2009} in an unmagnetized plasma except the factor $\gamma_v$, which was missing therein and in other several works. Such a factor appears due to consideration of the Lorentz transformations instead of the Galilean transformations in the stretched coordinates [Eq. \eqref{eq-stretch}]. While the Galilean transformation applies to nonrelativistic dynamics, the Lorentz transformation applies to wave dynamics in relativistic plasmas. We will show that such factor like $\gamma_v$ not only modifies the dispersion and nonlinear coefficients quantitatively but also modifies the instability domains, the instability growth rate, as well as the characteristics of solitons, especially when the group velocity $v_g$ is not much smaller than $c$. The latter can be justified when $k\lesssim1$ [see Eq. \eqref{eq-vg}]. The values of $k\ll1$ may not be acceptable because, otherwise, the EM wave will be dispersionless, which may correspond to low-frequency, long-wavelength phenomena, such as those described by the Korteweg-de Vries (KdV) equation. We also note that the first term in the nonlinear coefficient   $Q_{1}=\gamma_{v}/2\omega$ appears due to the nonlinear interactions of EM waves with the plasma density perturbation, while the second term  $Q_{2}=\gamma_{v}k^2\alpha^2/\omega\left(4\omega^2-1-4k^2\alpha\tilde{c_{s}}^2\right)$ is due to the nonlinear self-interaction of the carrier waves driven by the EM wave ponderomotive force and gets modified by the relativistic degenerate pressure (proportional to $\tilde{c_{s}}^2$).
%%%%%%
\section{Modulational Instability} \label{sec-MI}
Before we proceed to the instability analysis, it is noted that the NLS equation \eqref{eq-NLSE} admits a trivial plane wave time-dependent solution $\psi=\psi_{0}\exp{(iQ|\psi_{0}^2|\tau)}$, where  $\psi_{0}$ denotes a constant amplitude of the wave and $\Delta=-Q|\psi_{0}|^2$ is the nonlinear frequency shift.  Next, we modulate the EM wave amplitude and the phase against a plane wave perturbation (with the wave frequency $\Omega$ and the wave number $K$) by assuming $\psi=[\psi_{0}+\psi_{1}\cos(K\xi-\Omega\tau)]\exp[iQ|\psi_{0}|^2\tau+i\theta_{1}\cos(K\xi-\Omega\tau)]$. Substituting this expression of $\psi$ in Eq. \eqref{eq-NLSE} and separating the real and the imaginary parts, we obtain the following dispersion relation for the plane wave perturbations.
\begin{equation} \label{eq-disp-rel}
\Omega^2=(PK^2)^2\left(1-\frac{K_{c}^2}{K^2}\right),
\end{equation}
where $K_{c}=\sqrt{2|Q/P|}|\psi_{0}|$ is the critical value of the wave number of modulation $K$. From Eq. \eqref{eq-disp-rel},  it is clear that under the amplitude modulation, a plane wave solution of the NLS equation can be unstable against the plane wave perturbation   if $K<K_{c}$ or for wavelength values above the threshold, i.e., $\lambda_{c}=2\pi/K_{c}$ and $PQ>0$. In this case, the energy localization takes place induced by the nonlinearity to form a bright EM envelope soliton, i.e., a localized pulse-like envelope modulating the carrier wave.  The instability growth rate (replacing $\Omega$ by $i\Gamma$) is 
obtained as
\begin{equation} \label{eq-growthrate}
\Gamma=|P|K^2\sqrt{\frac{K_{c}^2}{K^2}-1}.
\end{equation}
Also, the maximum growth rate is attained at $K=K_{c}/\sqrt{2}$, i.e., $\Gamma_\text{max}=|Q||\psi_{0}|^2$, which explicitly depends on the nonlinear coefficient $Q$. 
On the other hand, when $PQ<0$,   Eq. \eqref{eq-disp-rel} shows that a plane wave form is stable under the modulation,  leading to the formation of a dark envelope  soliton,  which represents  a localized region of decreased amplitude. 
\par 
Before we proceed to the evolution of envelope solitons, we must carefully examine the sign of $PQ$ as it precisely determines whether the plane wave solution is stable or unstable under the amplitude modulation with plane wave perturbations.  We note that the dispersion coefficient $P$ is always positive, while the nonlinear coefficient $Q$ may be positive or negative depending on the carrier EM wave frequency $\omega$ or the wave number $k$ and the contribution from the relativistic degenerate pressure at finite temperature (proportional to $\tilde{c_s}^2$, which typically depends on the degeneracy parameter $\xi_e$ and the relativistic parameter $\beta_e$). In fact, $Q$ can be positive or negative according to when  $\omega \gtrless\omega_c$, where $\omega_c$ is the critical wave frequency, given by,
\begin{equation}
\omega_{c}=\sqrt{\frac{1}{2}\frac{2\alpha^2+4\alpha\nu_{e}\beta_{e}-1}{\alpha^2+2\alpha\nu_{e}\beta_{e}-2}}.
\end{equation} 
 \par
\begin{figure*}
\includegraphics[width=6in, height=3in]{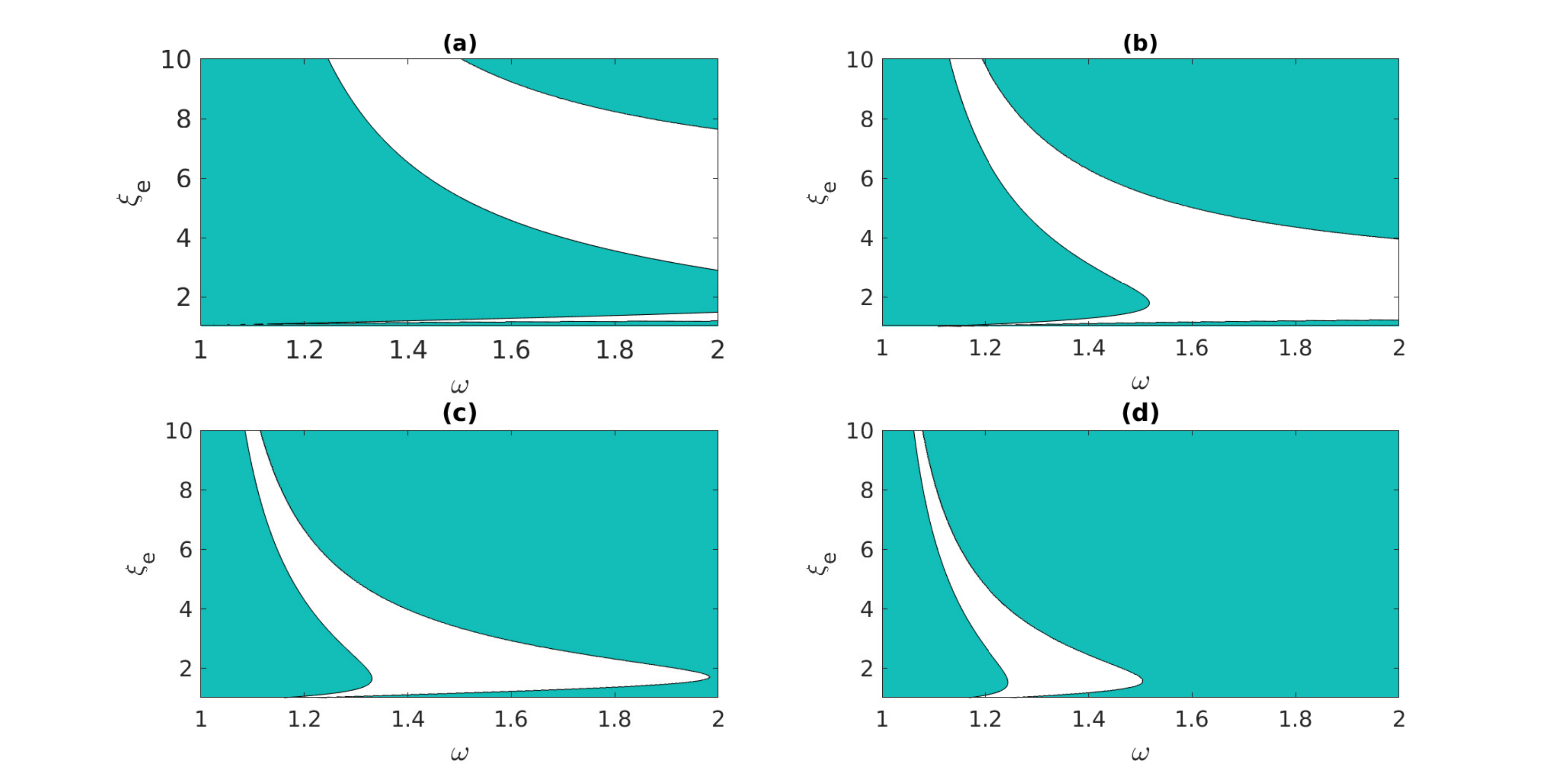}
\caption{The contour plots of $Q=0$ are shown in the $\omega\xi_e$ plane  for different values of $\beta_e<1$: (a) $\beta_e=0.3$,  (b) $\beta_e=0.5$,  (c) $\beta_e=0.7$, and (d) $\beta_e=0.9$.  The blank  (shaded)  region corresponds to the stable (unstable) region where $Q<0$ $(Q>0)$.  }
\label{fig1}
\end{figure*}
\begin{figure*}
\centering
\includegraphics[width=6in, height=3in]{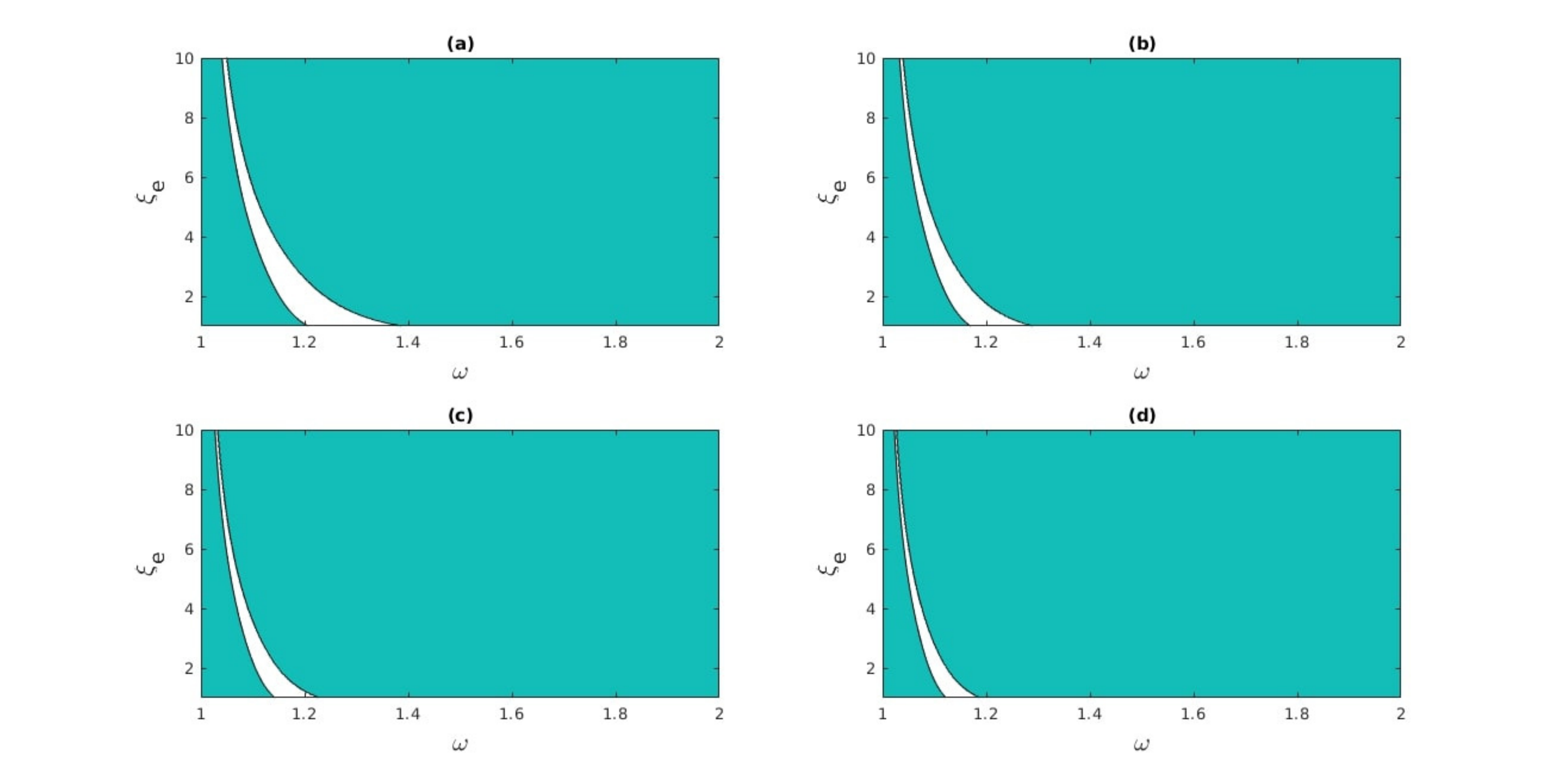}
\caption{ The same as in \ref{fig1}, but for different values of $\beta_e>1$: (a) $\beta_e=1.2$,  (b) $\beta_e=1.4$,  (c) $\beta_e=1.6$, and (d) $\beta_e=1.8$. }
\label{fig2}
\end{figure*}

 Typically, $\omega_c$ depends on the parameters $\xi_e$ and $\beta_e$. So, we numerically examine the sign of $Q$ in the $\omega\xi_e$-plane for different values of $\beta_e$, specifically when $\beta_e$ is slightly smaller and slightly larger than the unity. The results are shown in Figs. \ref{fig1}  and  \ref{fig2}, which  correspond  to the cases with $\beta_e<1$ and $\beta_e>1$ respectively. It is seen that the stable and unstable regions significantly depend on the two key parameters $\xi_e$ and $\beta_e$ in a particular domain of the EM wave frequency $\omega$. Figure \ref{fig1} shows that as $\beta_e$ increases from a fixed value $(<1)$ towards the unity, the stable (unstable) regions shrink (expand) and shift towards the unstable (stable) regions in the $\omega\xi_e$-plane. Consequently, as $\beta_e$ gradually increases and assumes values larger than the unity (Fig. \ref{fig2}), the stable region tends to shrink significantly, and we can see only the existence of the unstable region. It follows that while the lower values of $\beta_e~(<1)$ favor the modulational stable regions, its higher values $(>1)$ correspond to the instability. We note that in the absence of the degeneracy pressure \cite{borhanian2009}, $Q$ is always positive, implying modulational instability. Thus, we conclude that the modulated EM wave in an unmagnetized relativistic cold classical plasma and an unmagnetized relativistic degenerate plasma at finite temperature with a stronger influence of the thermal energy (than the rest mass energy) of electrons is always unstable. 
%%%%%%%%%%%%%%%%%% 
%%%%%%%%%%%%%%%%%%%%%%%%%%%%%%%%
\begin{figure*}
\centering
\includegraphics[width=6in, height=3in]{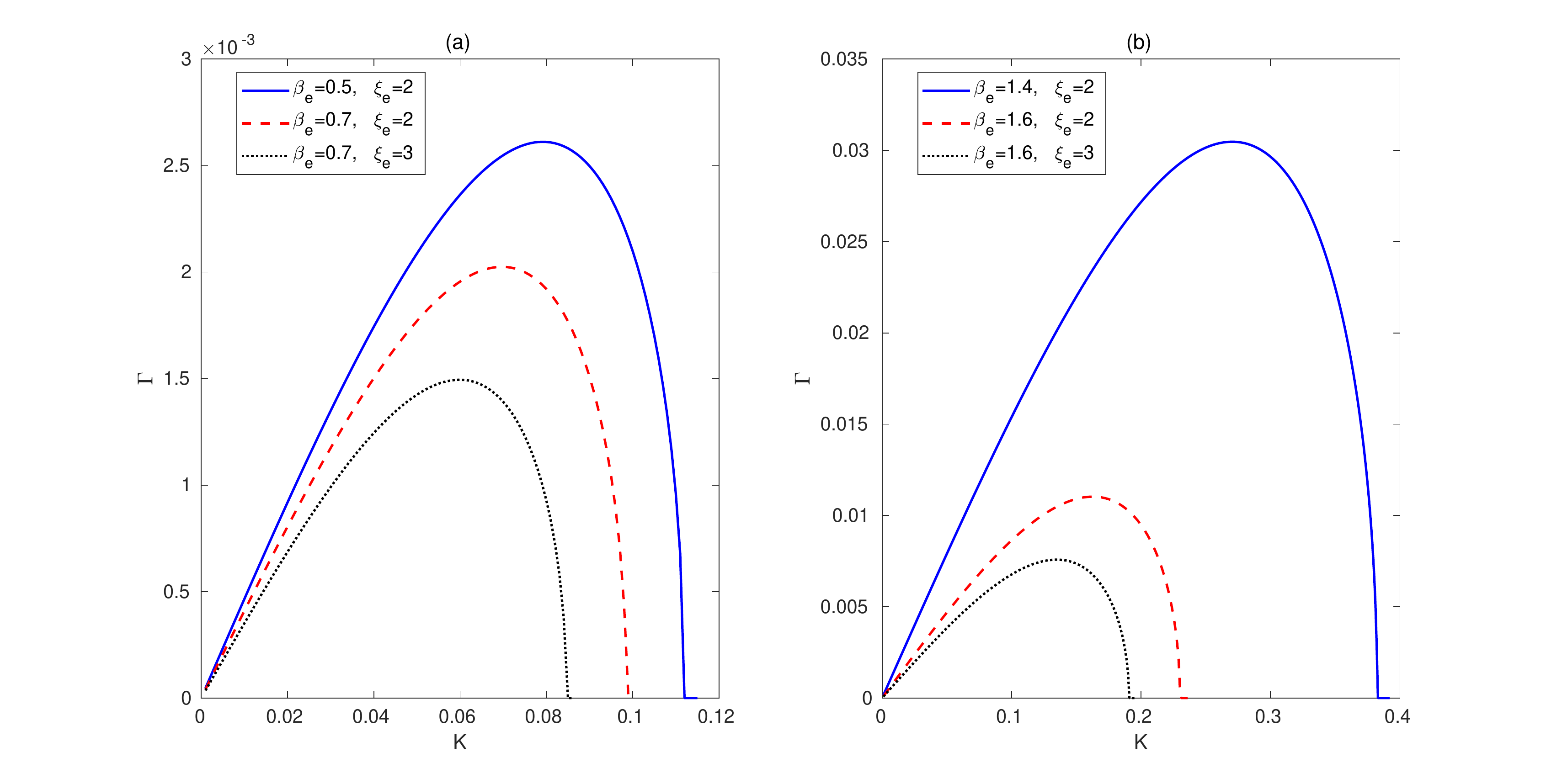}
\caption{The growth rate of instability $\Gamma$ is shown against the modulation wave number $K$ for a fixed $\omega=1.2$ and $\psi_{0}=0.07$ with different values of the parameters $\xi_{e}$ and $\beta_e$  [subplot (a) for $\beta_{e}<1$ and subplot (b) for $\beta_{e}>1$] as in the legends. }
\label{growthrate}
\end{figure*}
\par Having known the instability regions (shaded regions in Figs. \ref{fig1} and \ref{fig2}) in the planes of $\omega$ and $\xi_e$  for different values of $\beta_e$, we obtain the growth rate of instability $\Gamma$ against $K$ for a fixed $\omega=1.2$ and $a_{0}=0.1$, and for different values of the parameters $\beta_{e}$ and $\xi_{e}$. The results are displayed in Fig.\ref{growthrate}.  It is found  that in both the cases of $\beta_{e}<1$ [subplot (a)] and $\beta_{e}>1$ [subplot (b)], the growth rate decreases with increasing values of $\xi_{e}$ or $\beta_{e}$ (keeping the other parameter fixed), having cutoffs at lower values of the wave number of modulation. However, the maximum growth rate is relatively higher in the case of  $\beta_e>1$ than that with $\beta_e<1$, since in the former case, the nonlinear effect $(\sim |Q|)$ is more pronounced than the latter one.      
%%%%%%
%%%%
 \section{Envelope Solitons} \label{sec-EM-sol}
 Another interesting feature of Eq. \eqref{eq-NLSE} is that, apart from a plane wave solution, it also admits different localized  envelope soliton solutions,  which typically depend  on the sign of the product $PQ$. We note that  the total vector potential $a(z,t)$ can be represented as 
 \begin{equation}\label{eq-a-envelope}
 \begin{split}
  &a(z,t)\sim\psi(\xi,\tau)\exp[i(kz-\omega t)]+\text{c.c.},\\
  &\psi(\xi,\tau)=\psi_0(\xi,\tau)e^{i\theta(\xi,\tau)},
  \end{split}
 \end{equation}
 where the slowly varying wave envelope $\psi(\xi,\tau)$ with its slowly varying amplitude $\psi_0$ and the phase $\theta$ is determined by solving Eq. \eqref{eq-NLSE}.
 \subsection{Bright envelope soliton} 
\begin{figure*}
\centering
\includegraphics[width=6in, height=3in]{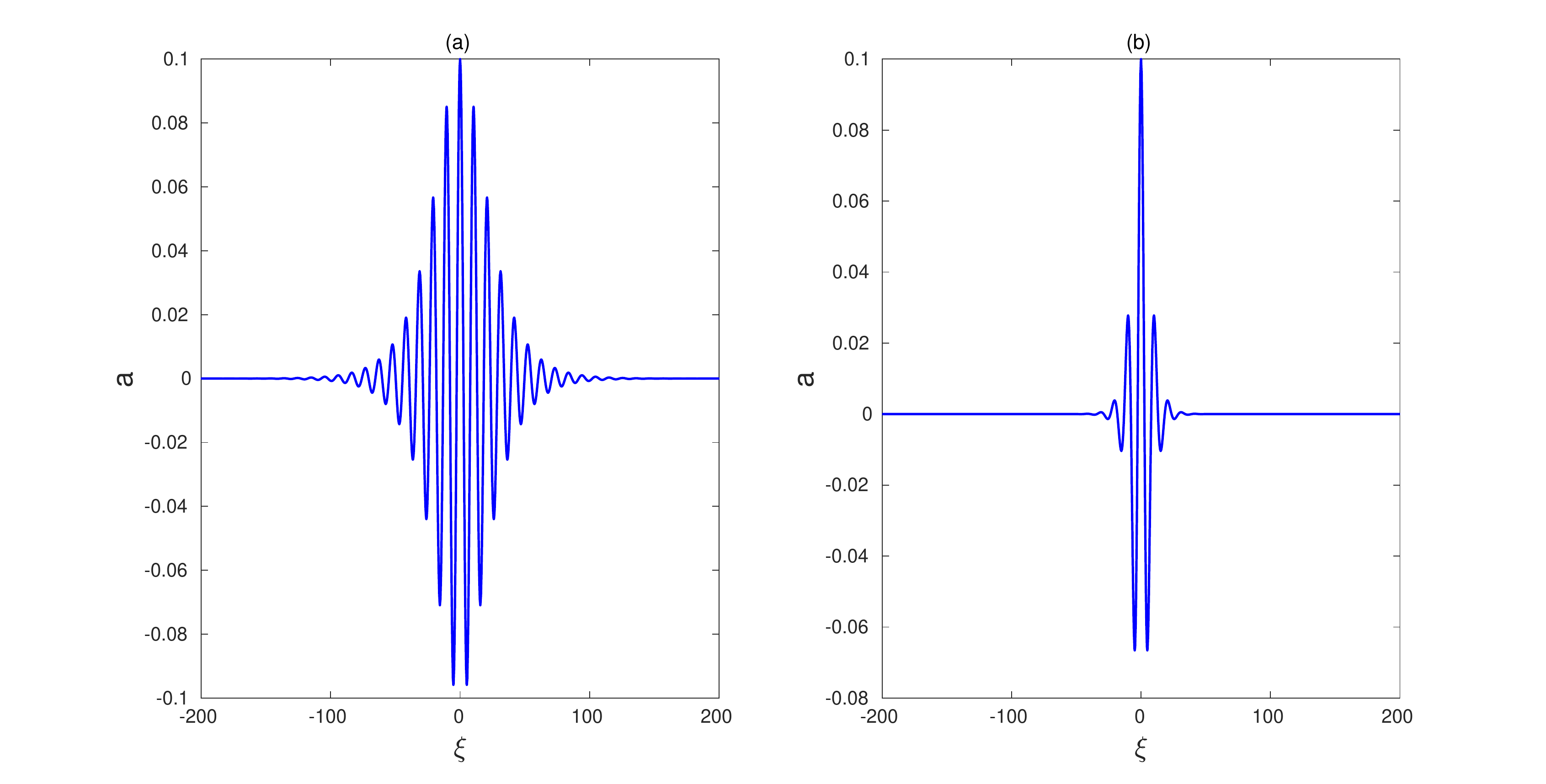}
\caption{A profile of the  bright EM envelope soliton is shown for (i) $\beta_e=0.5~(<1)$ [subplot (a)] and (ii) $\beta_e=1.4~(>1)$ [subplot (b)]. The fixed parameter values are  $\tau=0,~ U=0.5,~ \xi_{e}=2,~ \rho_{0}=0.06$, and $\omega=1.2$.}
\label{fig:bright-sol}
\end{figure*}

   For $PQ>0$, i.e., for $Q>0$ the wave is modulationally unstable, which leads to the formation of a bright EM envelope soliton, i.e., a localized pulse like envelope modulating the carrier wave. In this case, an exact analytic (bright soliton) solution of Eq. \eqref{eq-NLSE} can be obtained, which is given by \cite{fedele2002}
\begin{equation} \label{eq-bright-sol}
\begin{split}
&\psi_0=\rho_{0} \text{sech}\left(\frac{\xi-U\tau}{L} \right),\\
&\theta=\frac{1}{2P}\left[U\xi - \left(PQ\rho_0^2+\frac{1}{2}U^2\right)\tau\right].
\end{split}
\end{equation}
Here, $U$ is the constant speed  and  $L=\sqrt{2P/Q}/\rho_{0}$ is the spatial width of the pulse (traveling wave) such that $L\rho_0$ is a constant. A typical form of the bright envelope soliton is shown in Fig. \ref{fig:bright-sol} for $\beta_e<1$ [subplot (a)] and for $\beta_e>1$ [subplot (b)] for fixed values of the other parameters. We noted that as the thermal energy of electrons increases and exceeds the rest mass energy, the number of oscillations of the carrier wave forming the envelope gets significantly reduced. The localization occurs in a relatively shorter domain of the coordinate $\xi$.  Physically, as the values of the relativity parameter $\beta_e$ increase, the nonlinear coefficient $Q$ tends to become positive, thereby enhancing the self-focusing effect induced by the degenerate pressure, the relativistic effect, and the pondermotive force. The latter pushes electrons away from the region where the EM pulse is more intense, increasing the plasma refractive index and inducing a focusing effect. 
%%%%%%%%%%%%%%%%%%%%%%%%%%%%%%%
\subsection{Dark envelope soliton} 
When $PQ<0$, or, more precisely, $Q<0$, the plane wave is modulationally stable and may propagate in the form of a dark EM envelope soliton,  given by \cite{fedele2002},  
\begin{equation} \label{eq-dark-sol}
\begin{split}
&\psi_0=\rho_{0}\text{tanh}\left(\frac{\xi-U\tau}{L} \right),\\
&\theta=\frac{1}{2P}\left[U\xi +\left(PQ\rho_{0}^2-\frac{1}{2}U^2\right)\tau\right].
\end{split}
\end{equation}
%%%%%%%%%%%%%%%%%%%%%%%%%%%%%%%%%%%
\begin{figure*}
\centering
\includegraphics[width=6in, height=3in]{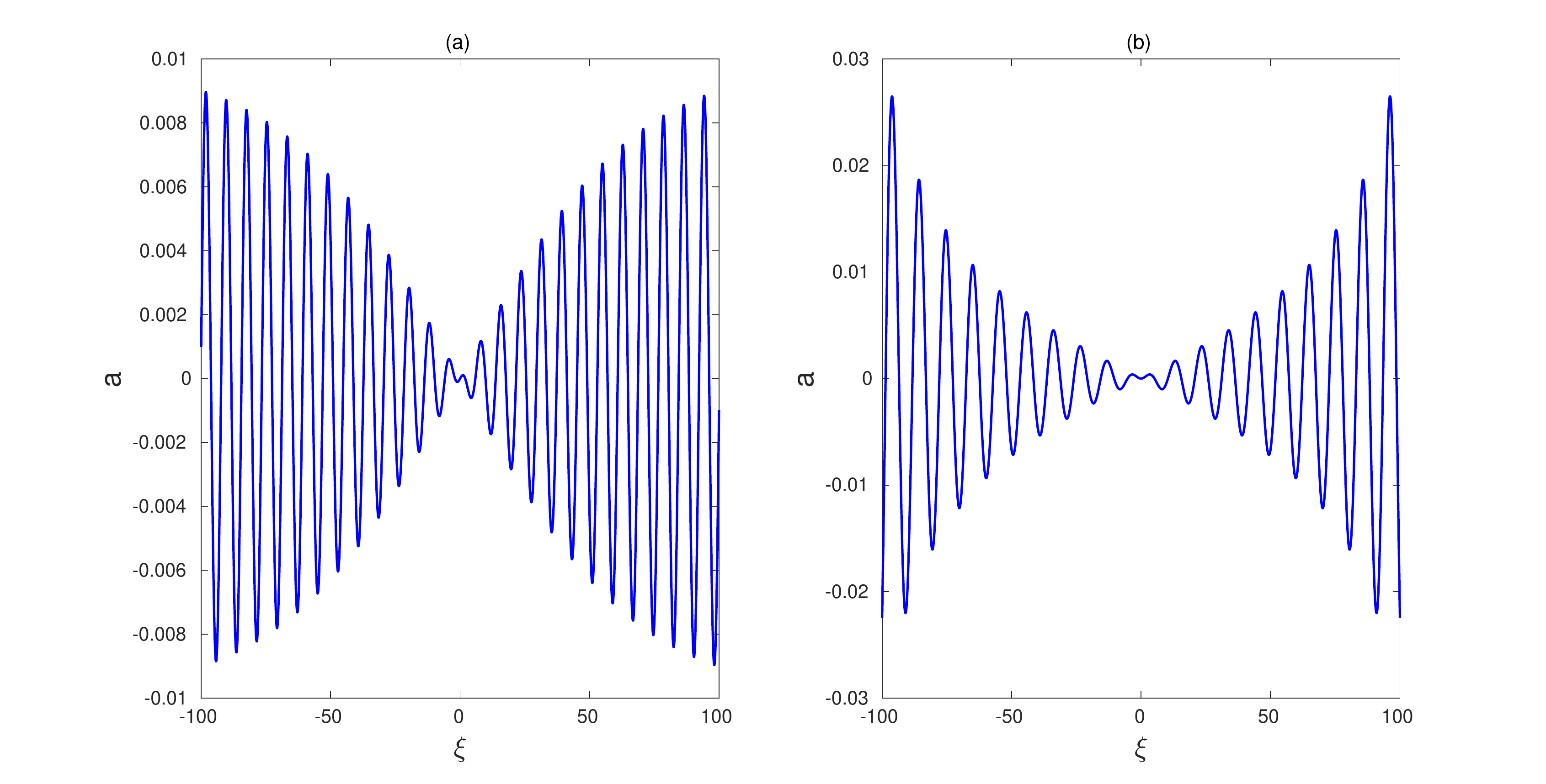}
\caption{A profile of the  dark EM envelope soliton is shown for (i) $\beta_e=0.5~(<1)$ [subplot (a)] and (ii) $\beta_e=1.2~(>1)$ [subplot (b)]. The fixed parameter values for subplots (a) and (b), respectively,  are  $(\tau=0,~ U=0.5,~ \xi_{e}=6,~ \rho_{0}=0.01,~\omega=1.6)$ and $(\tau=0,~ U=0.5,~ \xi_{e}=2,~ \rho_{0}=0.01,~\omega=1.21)$.}
\label{fig:dark-sol}
\end{figure*}
%%%%%%%%%%%%%%%%%%%%%%%%%%%%%%%%%%%%%
This soliton solution represents a localized region of a hole (void) traveling at a constant speed $U$ in a background that requires repulsive or defocusing nonlinearity. Also, the pulse width $L$ depends on the constant amplitude $\rho_{0}$ as $L=\sqrt{|2P/Q|}/\rho_{0}$. The profiles of the dark solitons are shown in Fig. \ref{fig:dark-sol} for $\beta_{e}<1$ [subplot (a)] and $\beta_{e}>1$ [subplot (b)]. We find that, in both cases, the amplitude approaches a zero value in the center of the pulse. Also, similar to the case of bright solitons, the number of oscillations of the carrier wave forming the envelope gets reduced in the case of $\beta_{e}>1$.
\par
It is important to note that the bright and dark envelope solitons are traveling wave solutions of the NLS equation \eqref{eq-NLSE}, associated with the modulational instability $(PQ>0)$ and stability $(PQ<0)$ of a plane waveform. However, when $PQ>0$, Eq. \eqref{eq-NLSE} can also admit other solutions, namely the Kuznetsov-Ma breather, the Akhmediev breather, and the Peregrine soliton. The latter is a limiting case for both the  Kuznetsov-Ma breather and the Akhmediev breather solitons. We illustrate these solitons in the following subsections \ref{sec-kuznetsov}-\ref{sec-peregrine}.  For more details about these solitons, see, e.g., a recent review work by Karjanto \cite{karjanto2021}. 
\subsection{Kuznetsov-Ma breather soliton} \label{sec-kuznetsov}
\begin{figure*}
\centering
\includegraphics[width=6in, height=3in]{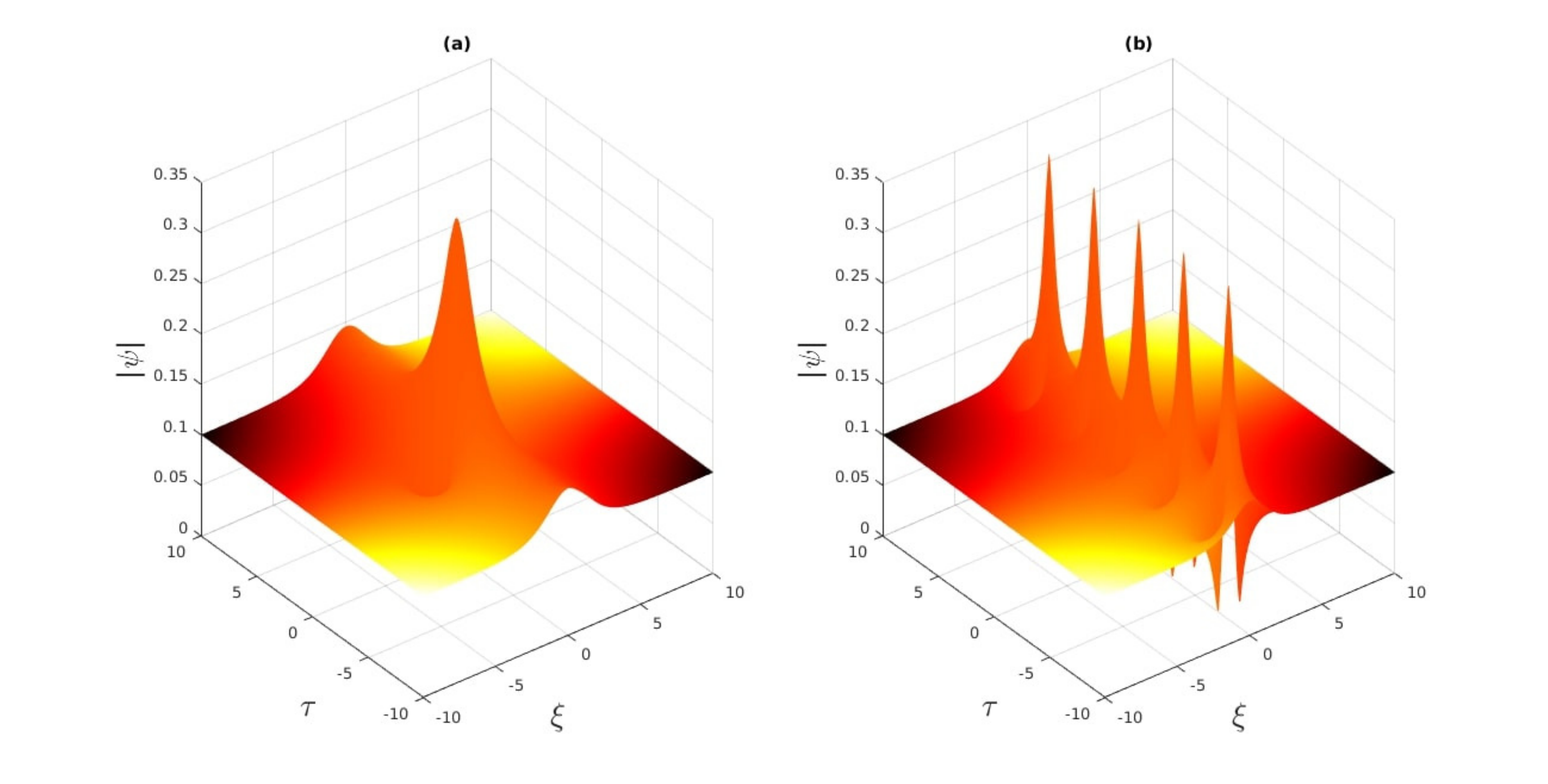}
\caption{Typical form of the Kuznetsov-Ma breather soliton [Eq. \eqref{eq-kuznetsov}] is shown for two different cases:  $\beta_e=0.5~(<1)$ [subplot (a)] and $\beta_e=1.7~(>1)$ [subplot (b)]. The fixed parameter values are $\omega=1.2$ and $\xi_e=2$.  }
\label{fig:kuznetsov}
\end{figure*}

For $PQ>0$, the NLS equation \eqref{eq-NLSE} has the following breather type soliton solution, called the  Kuznetsov-Ma breather. The latter is localized in the space variable $\xi$ but is periodic in the time variable $\tau$. 
\begin{equation}\label{eq-kuznetsov}
\psi=\left[1+\frac{\mu^3\cos\left(\frac{1}{2}Q\rho \tau\right)+i\mu\rho\sin\left(\frac{1}{2}Q\rho \tau\right)}
{2\mu\cos\left(\frac{1}{2}Q\rho \tau\right)-\rho\cosh\left(\mu\sqrt{\frac{Q}{2P}}\xi\right)}\right]e^{iQ\tau},
\end{equation}
where $\rho=\mu\sqrt{4+\mu^2}$. This expression of $\psi$ can be presented in an alternative form by considering $\mu=2\sin\phi$, so that $\rho=2\sin2\phi$.  It is important to note that the amplitude enhancement factor $(\sim1+\sqrt{4+\mu^2})$ of such breather solitons is more than three and it increases with increasing values of $\mu$. Furthermore, such breathers have been found to have potential applications as rogue waves or rogons in nonlinear dispersive media. Typical forms of these solitons are shown in Fig. \ref{fig:kuznetsov} for two different cases of $\beta_e<1$ [subplot (a)] and $\beta_e>1$ [subplot (b)]. It is interesting to note that as the electron thermal energy exceeds the rest mass energy (i.e., $\beta_e>1$), the nonlinearity enhances, which leads to the generation of multiple peaks (localization within a specified space interval) with  higher amplitudes but narrower widths.  
%%%%%%%%%%%%%%%%%%%%%%%%%%%%%%
\subsection{Akhmediev breather soliton} \label{sec-akhmediev}
\begin{figure*}
\centering
\includegraphics[width=6in, height=3in]{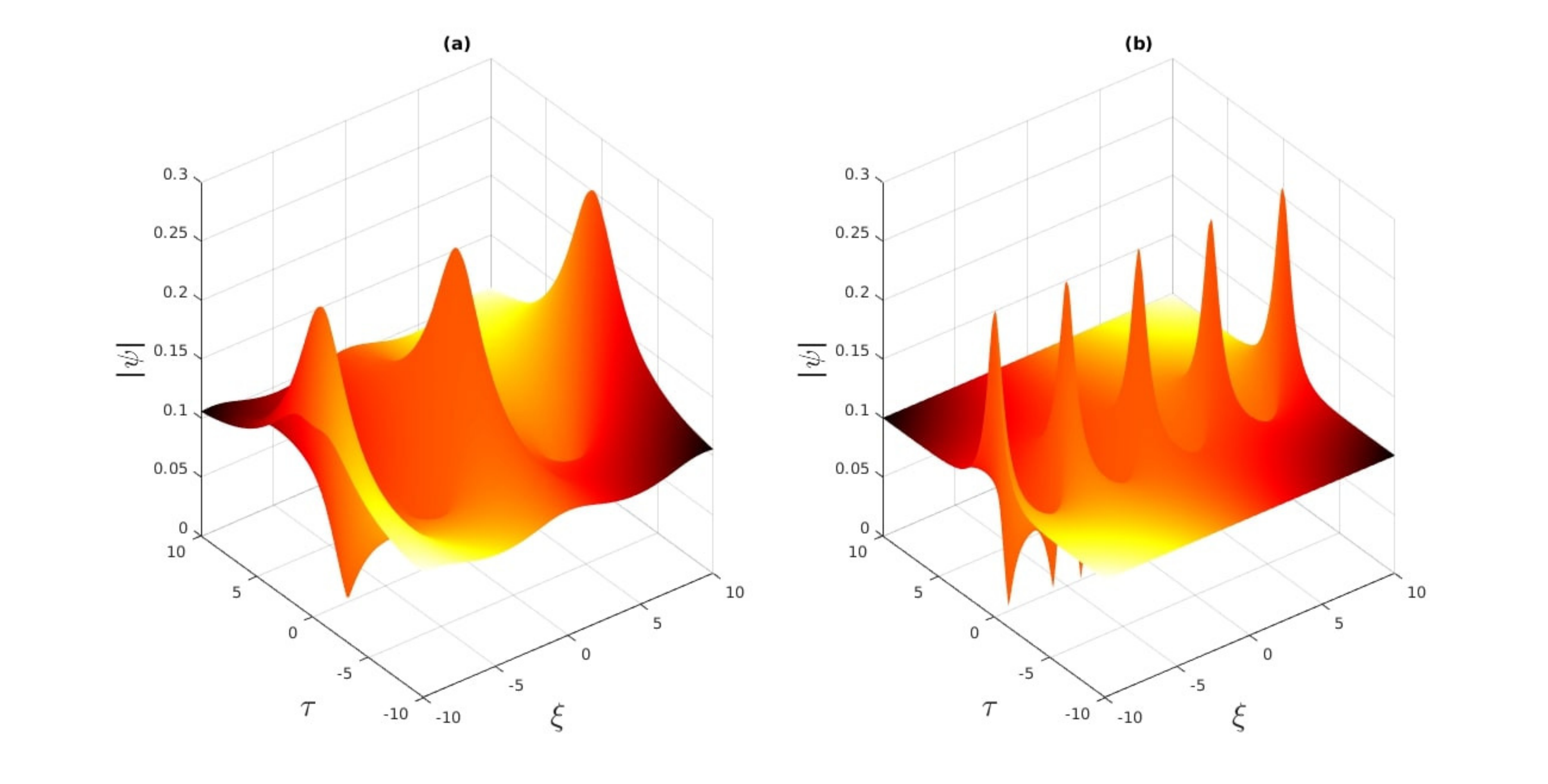}
\caption{Typical form of the Akhmediev breather soliton [Eq. \eqref{eq-akhmediev}] is shown for two different cases:  $\beta_e=0.5~(<1)$ [subplot (a)] and $\beta_e=1.7~(>1)$ [subplot (b)]. The fixed parameter values are $\omega=1.2$ and $\xi_e=2$.  }
\label{fig:akhmediev}
\end{figure*}

For $PQ>0$, the NLS equation \eqref{eq-NLSE} has the following breather type soliton solution, called the  Akhmediev breather, which is localized in the time variable $\tau$ but is periodic in the spatial coordinate $\xi$.
\begin{equation}\label{eq-akhmediev}
\psi=\left[1-\frac{\nu^3\cosh\left(\frac{1}{2}Q\sigma \tau\right)+i\nu\sigma\sin\left(\frac{1}{2}Q\sigma \tau\right)}
{2\nu\cosh\left(\frac{1}{2}Q\sigma \tau\right)-\sigma\cos\left(\nu\sqrt{\frac{Q}{2P}}\xi\right)}\right]e^{iQ\tau},
\end{equation}
where $\nu~(0\le\nu<2)$ and $\sigma$, respectively, stand  for a modulation frequency (or wave number) and the modulation growth rate such that $\sigma=\nu\sqrt{4-\nu^2}$. We note that in contrast to the Kuznetsov-Ma breather soliton,  the amplitude enhancement factor  $(\sim1+\sqrt{4-\nu^2})$ for the Akhmediev breather is below three and it decreases with increasing values of $\nu$. Also, similar to the Kuznetsov-Ma soliton,  the Akhmediev breather can act as one prototype rogue wave in which the modulational instability is considered as a possible mechanism for the energy localization.   Typical forms of these solitons are shown in Fig. \ref{fig:akhmediev} for two different cases of $\beta_e<1$ [subplot (a)] and $\beta_e>1$ [subplot (b)]. It is noted that as the electron thermal energy exceeds the rest mass energy (i.e., $\beta_e>1$), the nonlinearity enhances,   leading to the generation of multiple peaks (localization within a specified time interval) with  higher amplitudes but narrower widths.    
%%%%%%%%%%%%%%%%%%
%%%%%%%%%%%%%%%%%%%%%%%%%%%%%%
\subsection{Peregrine soliton} \label{sec-peregrine}
\begin{figure*}
\centering
\includegraphics[width=6in, height=3in]{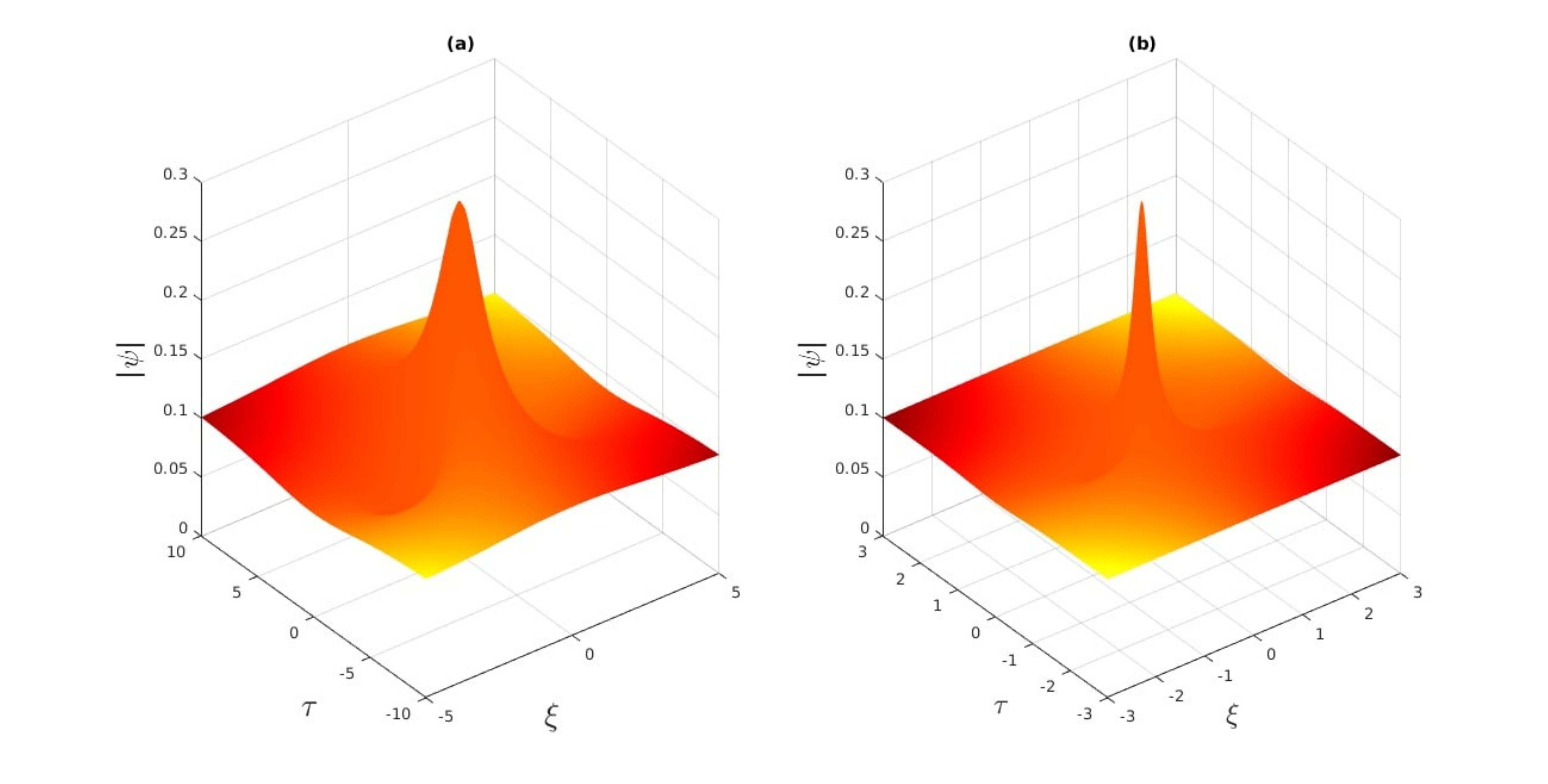}
\caption{ Typical form of the Peregrine soliton [Eq. \eqref{eq-peregrine}] is shown for two different cases:  $\beta_e=0.5~(<1)$ [subplot (a)] and $\beta_e=1.7~(>1)$ [subplot (b)]. The fixed parameter values are $\omega=1.2$ and $\xi_e=2$. }
\label{fig:peregrine}
\end{figure*}

Another kind of solution of the NLS equation \eqref{eq-NLSE} for  $PQ>0$ can also exist, which is localized in both the space and time variables $\xi$ and $\tau$. This is known as the Peregrine soliton or the rational solution, whose form is  written as
\begin{equation}\label{eq-peregrine}
\psi=\left[1-\frac{4(1+2iQ\tau)}{1+2(Q/P)\xi^2+4Q^2\tau^2}\right]e^{iQ\tau}.
\end{equation}
Like the Kuznetsov-Ma soliton and the Akhmediev breather, the Peregrine soliton is not a traveling wave. However, unlike them, it does not involve any free parameter. The amplitude amplification factor of the Peregrine soliton is precisely three, which can be obtained by taking limits of the two amplification factors of the Kuznetsov-Ma and the Akhmediev breathers as $\mu,~\nu\rightarrow0$. It has been shown that the Peregrine soliton acts as a  limiting behavior of the Kuznetsov-Ma and the Akhmediev breather solitons \cite{karjanto2021}. Although these two breather solitons are two prototypes of rogue waves,  the characteristics of Peregrine solitons are entirely consistent with those of rogue waves. They help explain the formation of those waves, which have a high amplitude and may appear from nowhere and disappear without a trace. Typical forms of the Peregrine soliton are shown in Fig. \ref{fig:peregrine} in two different cases of $\beta_e<1$ [subplot (a)] and $\beta_e>1$ [subplot (b)]. It is seen that as the value  of $\beta_e$ exceeds the unity, the soliton gets localized  in a small  space interval with a short duration of time. Although its amplitude remains almost unchanged, the width is decreased with increasing values of $\beta_e$.
%%%%%%%%%%%%
%%%%%%%%%
%%%%%%%
\section{Conclusion}
We have studied the modulational instability and the nonlinear evolution of slowly varying linearly polarized EM wave envelope in an unmagnetized relativistic degenerate plasma at finite temperatures. Specifically, we have focused on the regime where the thermal energy and the rest mass energy of electrons do not significantly differ, i.e., $\beta_e<1$ or $\beta_e>1$. However, the Fermi energy and the chemical potential energy are a bit higher than the rest mass energy of electrons. Starting from a set of relativistic fluid equations, the Fermi-Dirac pressure law, and the EM wave equation, and using the standard multiple-scale reductive perturbation technique, we have derived an NLS equation, which describes the evolution of slowly varying amplitude of EM wave envelopes. Then the modulational instability of a plane wave solution to the NLS equation is studied. The stable and unstable regions are obtained in the plane of the EM wave frequency $\omega$ and the normalized chemical potential $\xi_e$. We found that the parameter $\beta_e$ shifts the stable regions to unstable ones. When the thermal energy of electrons is almost double their rest mass energy (i.e., $\beta_e\sim2$), the plane wave solution is found completely unstable for a finite value of the carrier wave number $k$. However, the instability growth rate is lower at $\beta_e>1$, compared to $\beta_e<1$. The growth rate gets reduced at higher values of the normalized chemical potential $\xi_e$. 
\par
We have also shown that, in the modulational instability and stability regions, the slowly varying EM wave amplitude can evolve in the forms of localized bright and dark envelope solitons (traveling wave forms), respectively. We found that as the parameter $\beta_e$ tends to assume higher values, the localization of the EM envelope occurs in a smaller domain of space with a significant reduction of oscillations of the carrier wave forming the envelope. Furthermore, the formations of the Kuznetsov-Ma breather, the Akhmediev-breather, and the Peregrine solitons (other than traveling wave solutions), which can act as candidates for the evolution of EM rogue waves, are also shown.
\par 
Some important points concerning the strong field physics of laser-plasma interactions and the evolution of relativistic envelope solitons may be relevant to discuss. The present formulation applies to high-density degenerate (at finite temperature) plasmas, where the classical electrodynamics is still applicable for the interaction of electromagnetic fields with plasmas and the field strength ($\sim eE_\perp/m_e\omega c^2$) is well below the Schwinger critical field strength  ($\sim m_e^2c^3/e\hbar\approx1.32\times10^{18}$ V/m). That is, the theory is neither applicable to typical low-density classical plasmas (such as gaseous plasmas) nor to plasmas with pure quantum states. However, in critical or supercritical fields, various quantum electrodynamical or QED effects (e.g., photon-photon scattering, photon emission by a dressed lepton, and a photon decay into a dressed electron-positron pair)  will come into the picture that can lead to a variety of rich new phenomena  \cite{brodin2023,zhang2020}, not considered in the present study. Since the collective dynamics of  QED plasmas are significantly different from those of typical degenerate plasmas, reported here, the possibility of the emergence of Casimir-like effects in the formation of relativistic solitons and plasma density variations in strong fields (but well below the Schwinger limit) may be ruled out, because of quantum field fluctuations.  
\par 
Another important point is the formation of rogue waves in laser-plasma interactions. Although the origin of rogue waves is still a debatable issue, the modulational instability is considered as a possible mechanism for the energy localization both in space and time. We have seen that in addition to the localization, as the thermal energy of electrons starts increasing beyond their rest mass energy, the compression of pulses (with increasing amplitude) occurs due to the modification of the cubic nonlinearity associated with the relativistic EM wave driven ponderomotive force. Such intensification of laser pulses can cause an expulsion of plasmas, meaning that the QED effects could be realized in strong field laser-plasma interactions  \cite{shukla2005}.  
\par 
To conclude, the amplitude modulation of a plane wave solution of the NLS equation and the evolution of the slowly varying wave amplitude in the form of envelope solitons and rogue waves in relativistic degenerate plasmas at finite temperature should be helpful in laser fusion or laser-plasma interaction experiments, such as those, e.g., at the  NIF  \cite{hurricane2014} with particle number density approximately $10^{25}~\rm{cm}^{-3}$ or a bit higher.  
%%%%%%%%%%%%%%%%%%%%%%%
\section*{Acknowledgments} The authors thank all the three Referees for their insightful comments, which improved the manuscript in its present form.
%%%%%%%%%%%%%%%%%%%%%%%%%%%%% 
\section*{Author declarations}
\subsection*{Conflict of Interest}
The authors have no conflicts to disclose.
\subsection*{Author Contributions}
\textbf{Sima Roy:} Formal analysis (equal); Investigation (equal); Methodology (equal); Writing-original draft (equal). \textbf{Amar Misra:} Conceptualization (equal); Investigation (equal); Methodology (equal); Software (equal); Supervision (equal); Validation (equal); Writing-review \& editing (equal). \textbf{Alireza Abdikian:} Investigation (equal); Methodology (equal); Validation (equal).
%%%%%%%%%%%%%%%%%%%%%%%%%%
\section*{Data availability statement}
The data that support the findings of this study are available from the corresponding author upon reasonable request.
%%%%%%%%%%%%%%%%%%%%%%%%%5
\bibliographystyle{apsrev4-1} 
\bibliography{Reference}
\nopagebreak
\end{document}